\documentclass[prl,twocolumn,floatfix,altaffilletter,preprintnumbers,
               tightenlines,showpacs,showkeys]{revtex4-1}

\usepackage[colorlinks=true,citecolor=blue,linkcolor=blue]{hyperref}
\usepackage[normalem]{ulem}
\usepackage{amsmath,amssymb}
\usepackage{epsfig}
\usepackage{graphicx}               
\usepackage{url}
\usepackage{color}
\usepackage{multirow}
\usepackage{placeins}
\usepackage[dvipsnames]{xcolor}
\usepackage{epstopdf}

\usepackage{tikz}
\usetikzlibrary{trees}
\usetikzlibrary{decorations.pathmorphing}
\usetikzlibrary{decorations.markings}

\definecolor{jblue}  {RGB}{20,50,100}
\definecolor{npurple}  {RGB} {153, 51, 204}
\definecolor{wred}   {RGB}{217,0,56}
\definecolor{white}   {RGB}{255,255,255}

\definecolor{korange}   {RGB}{235, 80,  43}
\definecolor{korange2}   {RGB}{245, 100,  63}
\definecolor{kyelloworange}   {RGB}{255, 210,  110}
\definecolor{kyelloworange2}   {RGB}{240, 170,  90}
\definecolor{kred}   {RGB}{204,  102, 153}
\definecolor{kpurple}   {RGB}{153,  61, 190}
\definecolor{kpurplelight}   {RGB}{213,  161, 230}

\usepackage{cleveref}
\usepackage{subfigure} 
\usepackage{lipsum}
\definecolor{red}{rgb}{1.0, 0, 0}
 \usepackage{gensymb}
\allowdisplaybreaks

\bibpunct{[}{]}{,}{n}{}{,}

\newcommand{\ev}[1]{\ensuremath{\left\langle #1 %
                     \right\rangle}} 

\renewcommand{\vec}[1]{{\mathbf{#1}}}
\newcommand{\iso}[2]{{\ensuremath{{}^{#2}}\ensuremath{\rm #1}}}

\keywords{}

\begin{document}

\title{Dark Gamma Ray Bursts}

\author{Vedran Brdar\footnote{vbrdar@uni-mainz.de }}
\author{Joachim Kopp\footnote{jkopp@uni-mainz.de}}
\author{Jia Liu\footnote{liuj@uni-mainz.de}}
\affiliation{PRISMA Cluster of Excellence and
             Mainz Institute for Theoretical Physics,
             Johannes Gutenberg-Universit\"{a}t Mainz, 55099 Mainz, Germany}
\date{July 18, 2016}


\begin{abstract}
  Many theories of dark matter (DM) predict that DM particles can be captured
  by stars via scattering on ordinary matter. They subsequently condense into
  a DM core close to the center of the star and eventually annihilate.  In this
  work, we trace DM capture and annihilation rates throughout the life of a
  massive star and show that this evolution culminates in an intense
  annihilation burst coincident with the death of the star in a core collapse
  supernova. The reason is that, along with the stellar interior, also its DM
  core heats up and contracts, so that the DM density increases rapidly during
  the final stages of stellar evolution.  We argue that, counterintuitively,
  the annihilation burst is more intense if DM annihilation is a $p$-wave
  process than for $s$-wave annihilation because in the former case, more DM
  particles survive until the supernova.  If among the DM annihilation products
  are particles like dark photons that can escape the exploding star and decay
  to Standard Model particles later, the annihilation burst results in a flash
  of gamma rays accompanying the supernova.  For a galactic supernova, this
  ``dark gamma ray burst'' may be observable in CTA.
\end{abstract}

\maketitle


{\bf Introduction.}
A series of pioneering papers from the 1980s has established
that stars are likely to capture large amounts of dark matter (DM) particles
from the galactic halo~\cite{Silk:1985x, 
  Krauss:1985ks,      
  Srednicki:1986vj,
Gould:1987wi, Gould:1987ir, Gould:1987mp}.  This happens because DM particles
scattering on atomic nuclei or on previously captured DM particles inside the
star may lose enough energy to become gravitationally bound to it.
Subsequently, as they orbit its center, they experience additional scatterings
and eventually settle down into a thermalized DM core extending over
the inner regions of the star.  There, they can self-annihilate, and if some of
their annihilation products, in particular neutrinos, are able to leave the
star, observable signals are expected from the Sun
\cite{Silk:1985x,    
  Srednicki:1986vj,
  Gould:1987wi,
  Gould:1987ir,
  Gould:1987mp,
  Jungman:1995df,     
  Bruch:2009rp,       
  Nussinov:2009ft,    
  Menon:2009qj,       
  Kopp:2009et,        
  Batell:2009zp,      
  Blennow:2009ag,     
  Schuster:2009fc,    
  Meade:2009mu,       
  Bell:2011sn,        
  Lopes:2011rx,       
  Esmaili:2010wa,     
  Kearney:2012rf,     
  Esmaili:2012ut,     
  Arguelles:2012cf,   
  Bell:2012dk,        
  Bernal:2012qh,      
  Fukushima:2012sp,   
  Rott:2012qb,        
  Busoni:2013kaa,     
  Baratella:2013fya,  
  Ibarra:2014vya,     
  Blumenthal:2014cwa, 
  Danninger:2014xza,  
  Catena:2015uha,     
  Kumar:2015nja,      
  Blennow:2015xha,    
  Rott:2015nma,       
  Feng:2016ijc,       
  Vincent:2016dcp,    
  Murase:2016nwx}.    
Indeed, some of the strongest limits on DM scattering on nuclei are obtained
this way \cite{
Adrian-Martinez:2013ayv, 
Choi:2015ara,         
Aartsen:2016exj,      
Adrian-Martinez:2016ujo}. 
The potential impact of light DM particles on supernova cooling rates
has been studied in~\cite{Fayet:2006sa}.


The aim of this work is to follow the evolution of the DM core of a massive
($\gtrsim 8M_\odot$) star from the formation epoch all the way through its death
in a core collapse supernova.  As the star burns first hydrogen
and then heavier and heavier elements at its center~\cite{Kippenhahn:1990}, the
density and temperature of the baryonic matter increase, and as long as DM
scattering rates are sufficiently large, the DM core heats up and contracts in
the same way. As we will show in this paper, this leads to strongly enhanced DM
annihilation rates at the end of the stellar life cycle, and, in some DM
models, to observable signals. We dub these signals ``dark gamma ray bursts'',
(not to be confused with the scenario from
ref.~\cite{Banks:2014rsa}).


{\bf Dark Matter Capture.}
Assuming spherical symmetry, the rate at which DM particles
are captured by the star can be written as~\cite{Gould:1987ir, Busoni:2013kaa,
Baratella:2013fya, Taoso:2010tg}
\begin{align}
  C_\text{cap} = \sum_i \int_0^{R_\text{star}} \! dr \,
                   4 \pi r^2 \, \frac{dC_{i}(r)}{dV} \,.
  \label{eq:C-cap}
\end{align}
Here, the sum runs over the different isotopes in the star,
$R_\text{star}$ is its total radius, and
$dC_i(r)/dV$ is the capture rate per volume element~\cite{Gould:1987ir}. The
latter is proportional to the number density $n_i(r)$ of nuclei of type $i$, to
the galactic DM density $\rho_\text{DM}^\text{gal}$, and to the DM--nucleon
scattering cross section $\sigma_n$.

We take the radial density profile of the star and its chemical composition
from a simulation of a $12 M_\odot$ star by Heger et al.~\cite{Heger:2002cn,
Langer:1997bi, Woosley:2007as, Woosley:2006gw}.
$\rho_\text{DM}^\text{gal}$ depends on the distance
from the Galactic Center, and we model this dependence with an Einasto
profile~\cite{Einasto:1965, Graham:2005xx, Navarro:2008kc}.  For the
velocity distribution of galactic DM particles, we assume a Maxwell--Boltzmann
form with velocity dispersion $\bar{v} = 270 \,\text{km}\,\text{sec}^{-1}$.
This number is almost independent of the location of the star, except for stars very
close ($\ll 1$\,kpc) to the Galactic Center~\cite{Iocco:2015xga}.
The DM--nucleon scattering
cross section can include both spin-independent (SI) contributions
(for instance from scalar or vector interactions) and spin-dependent (SD)
contributions (for instance from axial vector interactions): $\sigma_n = \sigma_n^{\text{SI}} + \sigma_n^{\text{SD}}$.
We allow for both types of interactions and, unless otherwise noted, choose
$\sigma_n^{\text{SI}} = 10^{-46}\,\text{cm}^2$ and
$\sigma_n^{\text{SD}} = 10^{-40}\,\text{cm}^2$,
in accordance with constraints
\cite{Akerib:2015rjg, Amole:2015pla, Amole:2016pye}.
For these values, DM scattering on hydrogen is
entirely dominated by $\sigma_n^\text{SD}$, while for heavier nuclei,
$\sigma_n^\text{SI}$ is usually more relevant. The reason is that
most of the nuclei abundant in stars do not carry spin.
The exception is $\iso{N}{14}$, which has
an unpaired proton and an unpaired neutron. Lacking numerical values for
the corresponding spin matrix
elements $\ev{\vec{S}_p}$ and $\ev{\vec{S}_n}$, we approximate $\ev{\vec{S}_p}$
($\ev{\vec{S}_n}$) by its value for \iso{N}{15}
(\iso{C}{13})~\cite{Bednyakov:2004xq}.
For spin-independent scattering, nuclear form factors are modeled following
ref.~\cite{Gould:1987ir, Sivertsson:2009nx}.
We verified that the non-zero temperature of the nuclei
in the star is negligible as long as we consider only DM masses
$m_\text{DM} \gtrsim 10$~GeV~\cite{Busoni:2013kaa}.

In addition to scattering on nuclei, DM particles may also
scatter on previously captured DM particles if the underlying particle
physics model admits DM self-interactions~\cite{Zentner:2009is}.
Self-interacting DM is
motivated for instance by discrepancies between observed and predicted DM
distributions on small scales~\cite{Weinberg:2013aya}.
Though they are not crucial for our conclusions, we allow for self-interactions
mediated by a scalar or vector mediator of mass $M$ and with coupling
$\alpha' \equiv g'^2 / (4\pi)$.  The DM self-capture rate is computed in
analogy to the capture rate on ordinary matter~\cite{Gould:1987ir}.
We compute the DM--DM scattering cross section
$\sigma_\text{DM}$ in the classical limit, where $m_\text{DM} v / M \gg 1$ (with $v$
the DM velocity)~\cite{Khrapak:2003kjw,Khrapak:2014xqa},
and in the perturbative regime, where $\alpha' m_\text{DM} /
M \ll 1$~\cite{Feng:2016ijc,Tulin:2013teo}.  Between the two limiting regimes,
we use linear interpolation.  Numerically, we find that the self-capture rate
can be comparable to the capture rate on nuclei, especially when DM
annihilation is a $p$-wave process. As we will see below, the number of DM
particles in the stellar core is larger in this case.

After the initial DM--nucleon or DM--DM scattering that captures a DM particle
by reducing its velocity below the escape velocity,
the particle scatters multiple
more times, eventually thermalizing with the baryonic matter. The DM
core of the star can therefore be assigned a temperature $T_\text{DM}$,
which we take equal to the baryon temperature at its average
radius~\cite{Busoni:2013kaa},
\begin{align}
  \bar{r} = \frac{\int_0^{R_\text{star}}\!d^3r \, r \, n_\text{DM}(r)}
                 {\int_{0}^{R_{\text{star}}}\!d^3r \, n_\text{DM}(r)} \,.
\end{align}
The DM number density is parameterized as~\cite{Busoni:2013kaa}
\begin{align}
  n_\text{DM}(r) = n_0 \, \exp[-m_{\text{DM}} \phi(r) / T_\text{DM}] \,.
  \label{eq:nDM}
\end{align}
with $\phi(r)$ the gravitational potential at radius $r$.

Scattering of a previously captured DM particles on nuclei or other DM
particles can also increase its energy and lead to
evaporation~\cite{Gould:1987wi, Gould:1987mp, Busoni:2013kaa, Chen:2014oaa},
but we have confirmed numerically that for the DM masses of interest to us
($m_\text{DM} \gtrsim 10 \text{GeV}$), this effect is negligible.


{\bf Dark matter annihilation.}
The DM annihilation rate is given by \cite{Busoni:2013kaa,Baratella:2013fya}
\begin{align}
  \Gamma_\text{ann}
    =      \frac{1}{2} \int\!d^3r \, \ev{\sigma v_\text{rel}} n_\text{DM}^2(r)
    \equiv \frac{1}{2} C_\text{ann} N_\text{DM}^2 \,.
  \label{eq:Gamma-ann}
\end{align}
Here, $\ev{\sigma v_\text{rel}}$ is the annihilation cross section, multiplied
by the relative velocity $v_\text{rel}$ of the annihilating DM particles
and averaged over their thermal velocity distribution.
We have here defined the
annihilation coefficient $C_\text{ann}$ in terms of the total number of DM
particles in the stellar core, $N_\text{DM} = \int\!d^3r \, n_\text{DM}(r)$.

Below, we distinguish between models in which $\ev{\sigma v_\text{rel}}$ is
approximately independent of $v_\text{rel}$ ($s$-wave annihilation) and models
where $\ev{\sigma v_\text{rel}} \propto \ev{v_\text{rel}^2}$ ($p$-wave
annihilation)~\cite{Tulin:2013teo, Liu:2014cma}. As a benchmark model for $s$-wave
($p$-wave) annihilation, we consider DM annihilation into dark photons $A'$
(dark sector scalars $\phi$) through $t$-channel and $u$-channel diagrams.  We
choose the annihilation cross section such that $\ev{\sigma v_\text{rel}} = 4.4
\times 10^{-26}\,\text{cm}^3/\text{sec}$ at $\ev{v_\text{rel}^2} \simeq 0.24$
\cite{Fox:2011fx}, characteristic for thermally produced Dirac DM particles
\cite{Steigman:2012nb, Ibarra:2015fqa}. In the star, at temperatures much lower
than in the early Universe, the annihilation cross section is then much
smaller for $p$-wave annihilation than for $s$-wave annihilation.  Therefore,
more DM can accumulate inside the star in the $p$-wave case.  If $A'$ or $\phi$
is much lighter than $m_\text{DM}$, also Sommerfeld enhancement needs to be
taken into account~\cite{Sommerfeld, Cassel:2009wt, Tulin:2013teo}.


\begin{figure}
  \vspace*{-0.3cm}
  \hspace*{-0.4cm}
  \begin{tabular}{c@{}c}
    \includegraphics[width=0.54\columnwidth]{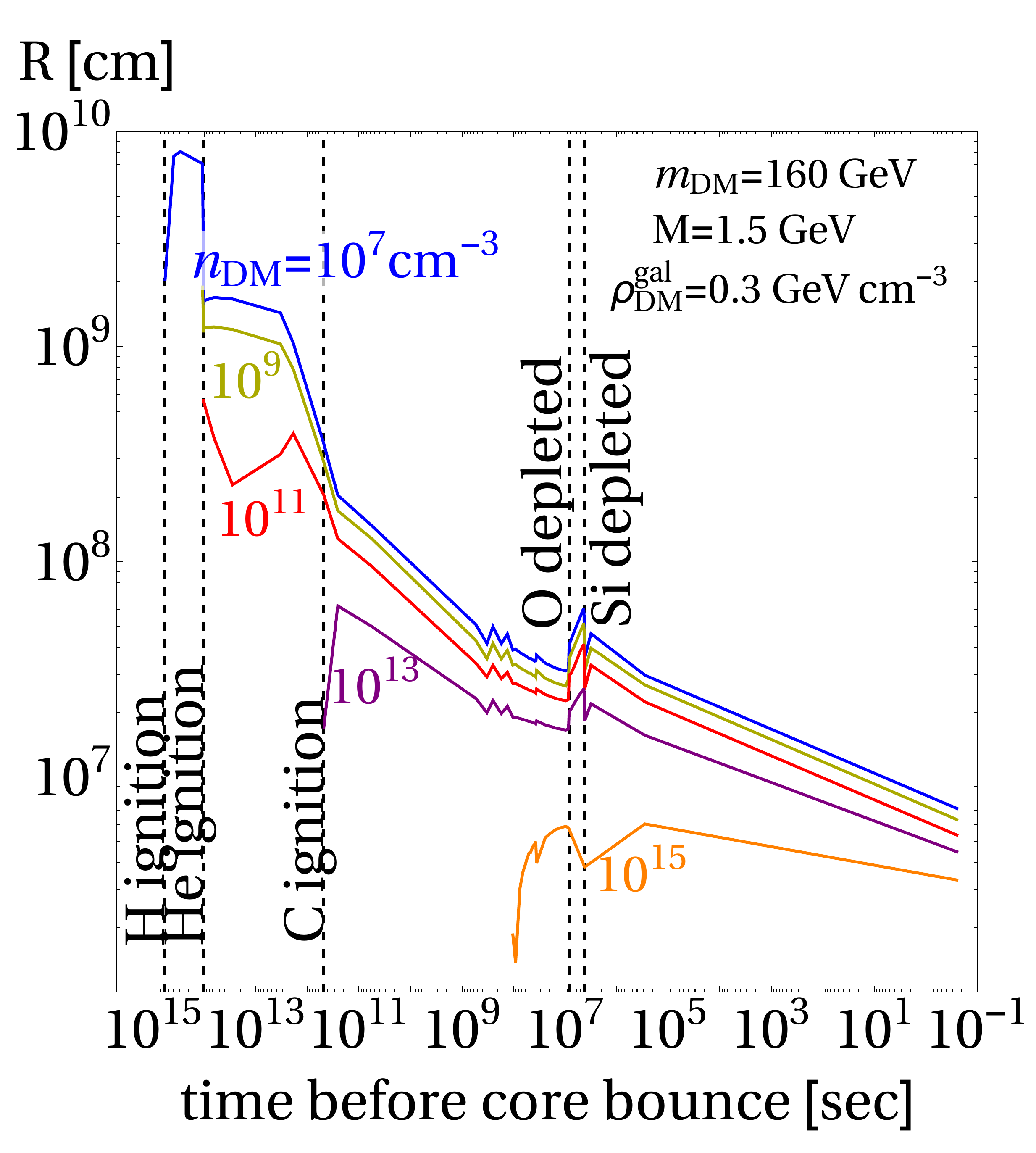} &
    \includegraphics[width=0.54\columnwidth]{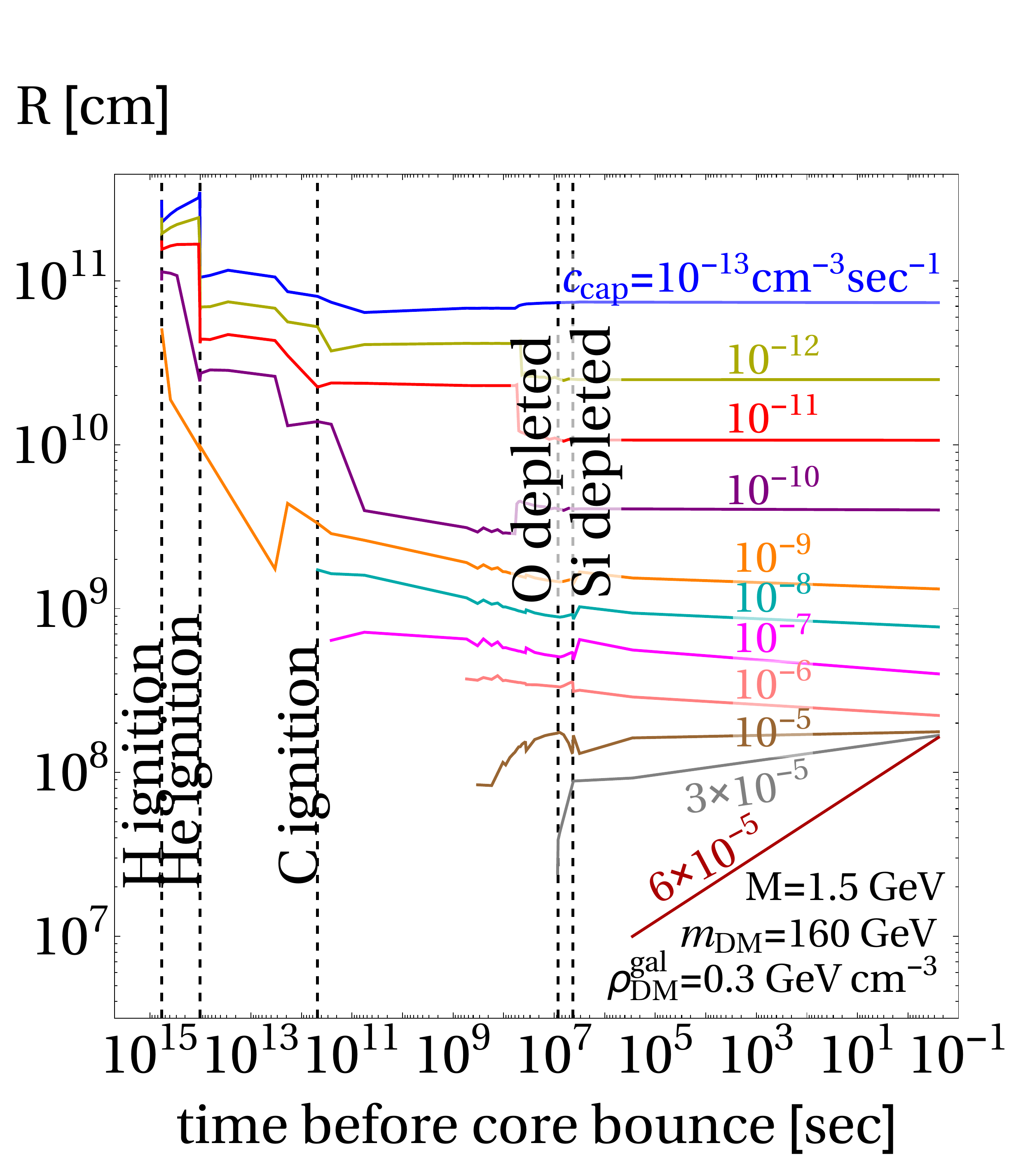} \\
    (a) & (b)
  \end{tabular}
  \caption{Contours of constant DM number density $n_\text{DM}$ (left) and constant
    DM capture rate $c_\text{cap}$ (right) in the $t$--$R$ (time--radius) plane.
    $c_\text{cap}$ is defined as $\sum_i dC_i(r)/dV$ where the sum runs over different nuclei.
    Note that the time axis is reversed to better
    reflect the fast evolution in the late stages.  We have assumed $s$-wave
    annihilation of 160\,GeV DM particles into dark 
    photons of mass $M = 1.5$\,GeV.
  }
  \label{fig:t-vs-r}
\end{figure}

{\bf Evolution of the DM population inside the star.}
The number of DM particles gravitationally bound to the star obeys the
differential equation
\begin{align}
  \dot{N}_\text{DM}(t) = C_\text{cap}(t) - C_\text{ann}(t) N_\text{DM}(t)^2
                             + C_\text{self}(t) N_\text{DM}(t) \,,
  \label{eq:dNdt}
\end{align}
with $C_\text{cap}$ and $C_\text{ann}$ from \cref{eq:C-cap,eq:Gamma-ann},
respectively, and with $C_\text{self}$ the self-capture coefficient
\cite{Zentner:2009is}. The time dependence of $C_\text{cap}$, $C_\text{ann}$
and $C_\text{self}$ arises from the time-dependent density, temperature, and
chemical composition of the baryonic matter.  Our input data for these
quantities from ref.~\cite{Heger:2002cn} consists of 52 non-linearly spaced
snapshots between the beginning of hydrogen burning in the stellar core and the
pre-supernova stage, 250~ms before core bounce.  Between snapshots, we
interpolate linearly.  We assume quasi-instantaneous thermalization of captured
DM particles because during most of the star's life, the time between
DM--nucleus collisions is much shorter than the time scales over which the
properties of the baryonic matter change significantly.  
Note that DM particles captured during the last $\sim
1000$~yrs do not thermalize, but their number is negligibly small.
Particles that were captured earlier and are already thermalized
follow the evolution of the baryonic matter even shortly before the supernova.
For instance, in the
hot iron core just before collapse (temperature $\sim 3$\,MeV, density
$10^{14}\,\text{grams}/ \text{cm}^3$), the collision time of a
100\,GeV DM particle via $\sigma^\text{SI}_n$ is of order 0.001\,sec.

The evolution of the stellar DM core is illustrated in \cref{fig:t-vs-r}. We
see that the DM core grows during hydrogen burning
and then contracts along with the baryonic matter.
The reason is that capture is very efficient at early times
(see \cref{fig:t-vs-r} (b)), thanks
to the large spin-dependent scattering rate on hydrogen. Later, when hydrogen
is depleted, the capture rate drops.

\begin{figure}
  \hspace*{-0.4cm}
  \begin{tabular}{cc}
    \includegraphics[width=0.52\columnwidth]{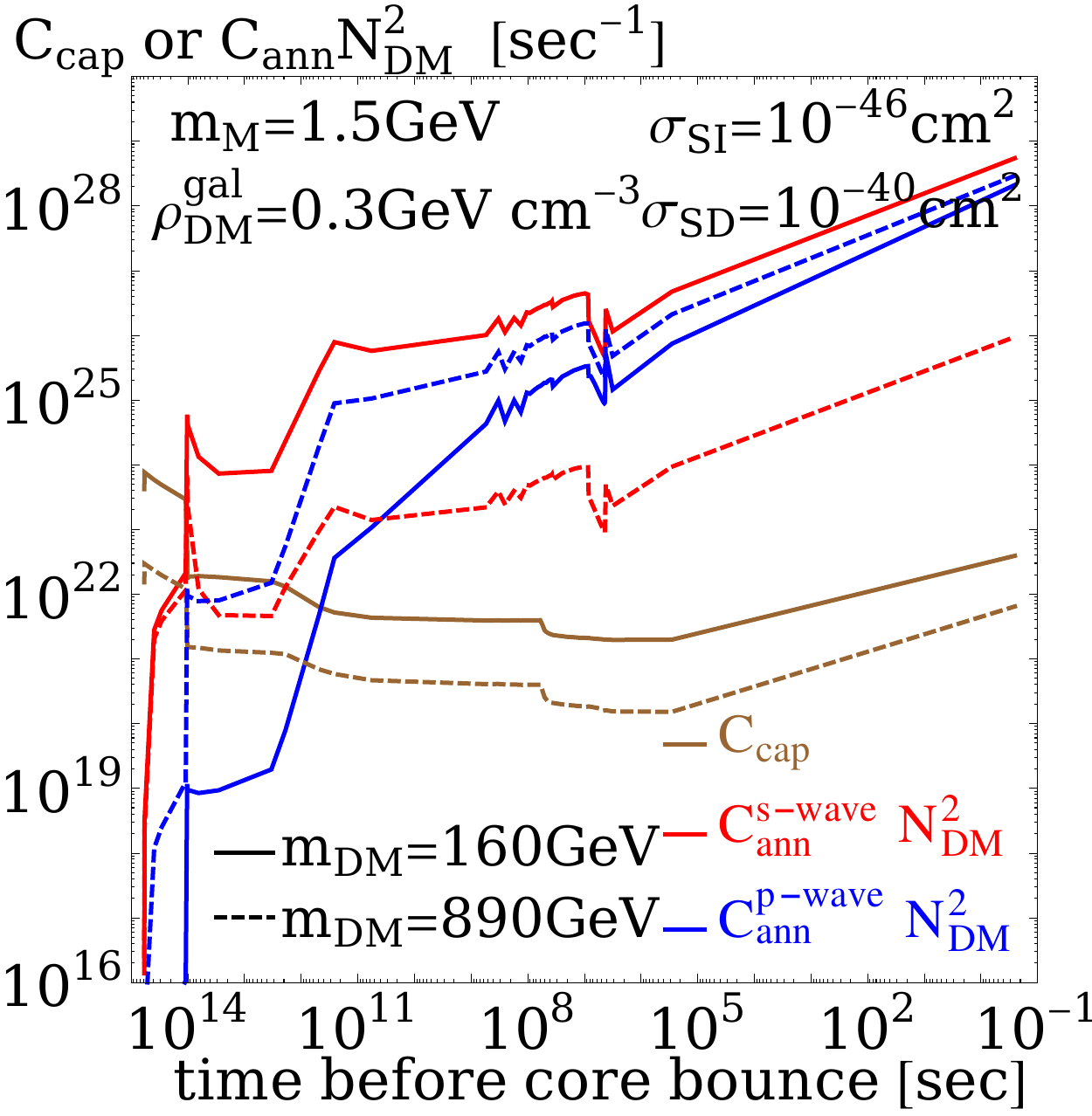} &
    \includegraphics[width=0.52\columnwidth]{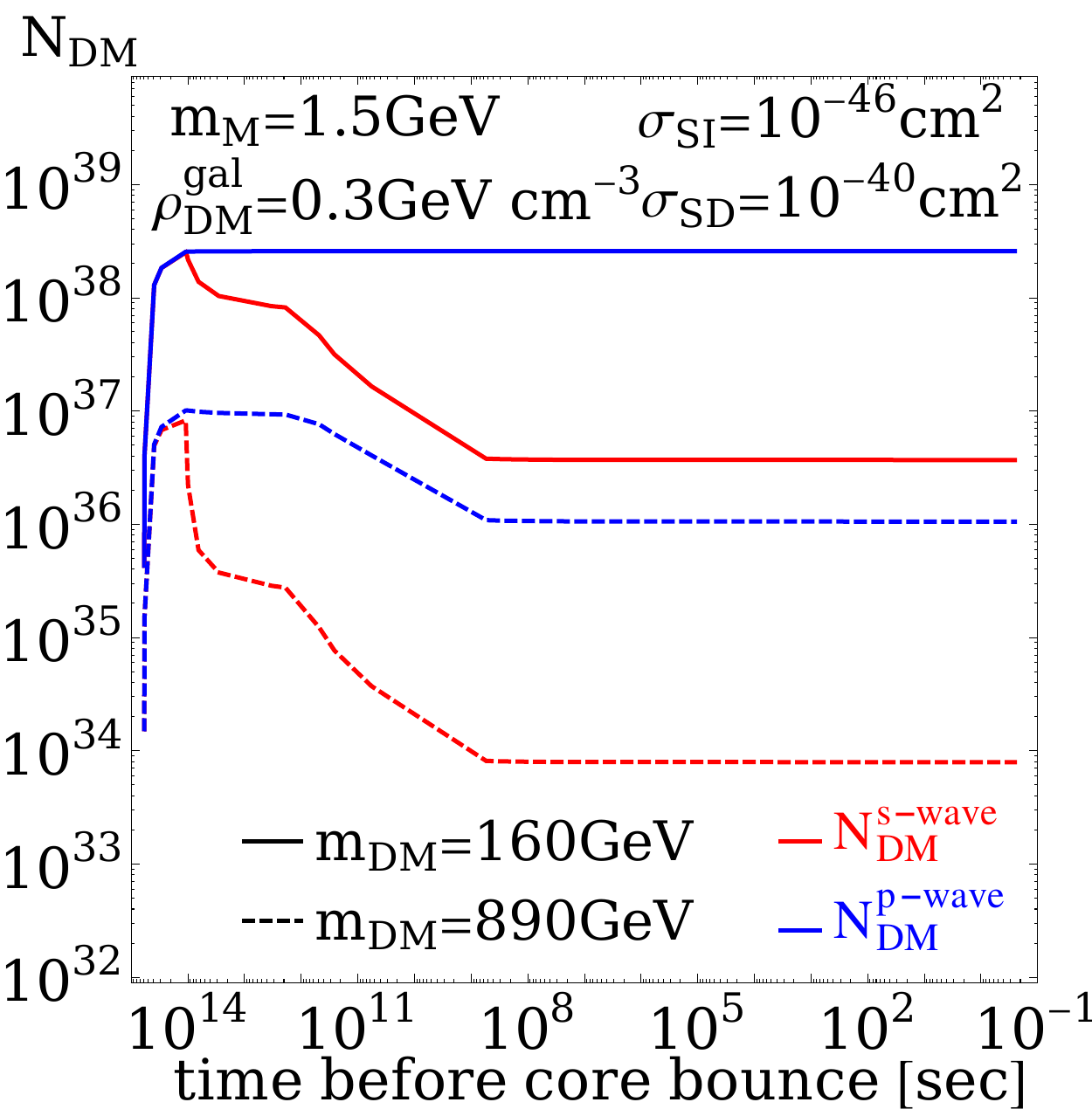} \\
    (a) & (b)
  \end{tabular}
  \caption{
    (a) Time evolution of the radius-integrated capture and annihilation rates
    (see \cref{eq:C-cap,eq:Gamma-ann}) for two different models ($s$-channel annihilation
    to dark photons and $p$-channel annihilation to dark scalars) and for two
    different DM masses.
    (b) Time evolution of the total number of captured DM particles $N_\text{DM}$ for
    the same benchmark models.
  }
  \label{fig:t-evolution}
\end{figure}

The time evolution of several radius-integrated parameters of the DM
core is shown in \cref{fig:t-evolution}.  Panel (a) illustrates again that
the capture rate rapidly decreases
after core hydrogen burning ceases.
This decrease is accompanied by
an increase in the DM annihilation rate, caused by contraction of the DM core.
(The slight increase of $C_\text{cap}$
shortly before the supernova is due to contraction of residual hydrogen.)
For $s$-wave annihilation (red curves), the number $N_\text{DM}$ of DM particles
in the star
decreases by several orders of magnitude once core hydrogen burning is over,
see \cref{fig:t-evolution} (b).  For velocity-suppressed $p$-wave annihilation
(blue curves), the qualitative behavior of the
annihilation rate is similar, but since its magnitude is much smaller,
$N_\text{DM}$ remains larger.

Increasing $m_\text{DM}$ leads to a moderate decrease in
$C_\text{cap}$.  For $s$-channel annihilation it also leads to a significant
decrease of the annihilation rate because of the smaller $N_\text{DM}$.
For $p$-wave annihilation, the
annihilation rate \emph{increases} with $m_\text{DM}$ because of more efficient
Sommerfeld enhancement.

It is interesting to observe from \cref{fig:t-evolution} (a) that, in contrast to
the more widely studied DM core of the Sun, the DM core of a massive star
evolves so fast that capture and annihilation never reach a stable equilibrium.
$C_\text{cap}$ is larger during hydrogen burning, while the annihilation rate
becomes dominant later.
The cross-over between the two occurs much later for $p$-wave annihilation
than for $s$-wave models.

We briefly comment on the dependence of our results on the DM scattering and
annihilation cross sections.  If $\sigma_n$ (or $\rho_\text{DM}^\text{gal}$) is lowered
by a factor 10--100, the number of captured DM particles $N_\text{DM}(t)$ decreases
correspondingly if DM annihilation is a $p$-wave process, but only slightly for
$s$-wave annihilation. The reason is that already a slight decrease in $N_\text{DM}(t)$
leads to an appreciable decrease in the annihilation rate, which counteracts
the smaller capture rate.  This feedback effect is relevant mostly in the
$s$-wave case, while for $p$-wave annihilation, it becomes appreciable only for
heavy $m_\text{DM}$, where Sommerfeld enhancement is strong.  For the same reason, increasing
(decreasing) the DM annihilation rate by a factor of 10 in $s$-wave models
leads to a decrease (increase) of $N_\text{DM}(t)$ by a similar factor, while doing the
same in $p$-wave scenarios has only a small impact on $N_\text{DM}(t)$.  Annihilation
cross sections larger than the canonical value $4.4 \times
10^{-26}\,\text{cm}^3 /\text{sec}$ are possible for example in models of
multi-component DM~\cite{Boehm:2003ha,Hur:2007ur,Dienes:2014via}, while
annihilation cross sections smaller than the canonical value are required in
scenarios with late time entropy production~\cite{Baltz:2001rq,
  Bezrukov:2009th, Hooper:2013nia}.


{\bf The annihilation burst.}
We see from \cref{fig:t-evolution} (a) that the DM annihilation rate increases
sharply just prior to the supernova.  This trend continues when the iron core
of the star collapses and eventually reaches nuclear density $\rho_\text{SN}
\sim 10^{14}\,\text{g} / \text{cm}^3$.  The result is a sudden burst of DM
annihilations. To compute the annihilation rate during this burst, we assume
that DM particles located within the newly forming neutron star ($R_\text{core}
\simeq 30$\,km) \cite{Woosley:2006ie} thermalize instantaneously with the
baryonic matter. This is justified because in the proto-neutron star,
collisions via $\sigma_n^\text{SD}$ occur every $\sim 10^{-7}$\,sec.
To be conservative, DM particles at larger radii are discarded
at this stage.  We assume for simplicity that throughout the supernova and the
early cooling stage of the newborn neutron star (first $\sim 10^3$\,seconds),
its temperature is constant at $T_\text{SN} \sim 3$\,MeV.  Since according to
\cref{eq:nDM}, cooling leads to further contraction of the DM distribution and
thus to more efficient annihilation, this assumption is conservative.  The
annihilation rate during the burst is given by \cref{eq:Gamma-ann}, where now
\cite{Baratella:2013fya}
\begin{align}
  C_\text{ann}^\text{SN}
    = \ev{\sigma v_\text{rel}} \bigg( \frac{G_N m_\text{DM} \rho_\text{SN}}
                                           {3 T_\text{SN}} \bigg)^{3/2} \,
  \label{eq:C-ann-SN}
\end{align}
and $N_\text{DM}(t)$ evolves as
\begin{align}
  N_\text{DM}(t) = \frac{N_{0}}{1 + t \, C_\text{ann}^{SN} N_{0}} \,
  \label{eq:N-t-ann-only}
\end{align}
(see \cref{eq:dNdt}, neglecting the capture terms).  Here, $N_0$ is the number
of DM particles in the star just prior to the supernova.
\Cref{eq:N-t-ann-only} also determines the duration of the DM annihilation
burst $\Delta t_\text{burst} \sim (C_\text{ann}^{SN} N_{0})^{-1}$.  For
$s$-wave annihilation, $\Delta t_\text{burst}$ varies between
$10^2$--$10^3$\,sec for the DM masses and cross sections considered in this
paper, while for $p$-wave annihilation, $\Delta t_\text{dur}$ can be as small
as 10~sec or as large as $10^4$\,sec.
The reason is the weaker dependence of $N_0$ on the model parameters for
$s$-wave annihilation, caused by the stronger feedback of annihilation
on $N_\text{DM}(t)$ throughout the star's life.

\begin{figure}
  \centering
  \includegraphics[width=0.7\columnwidth]{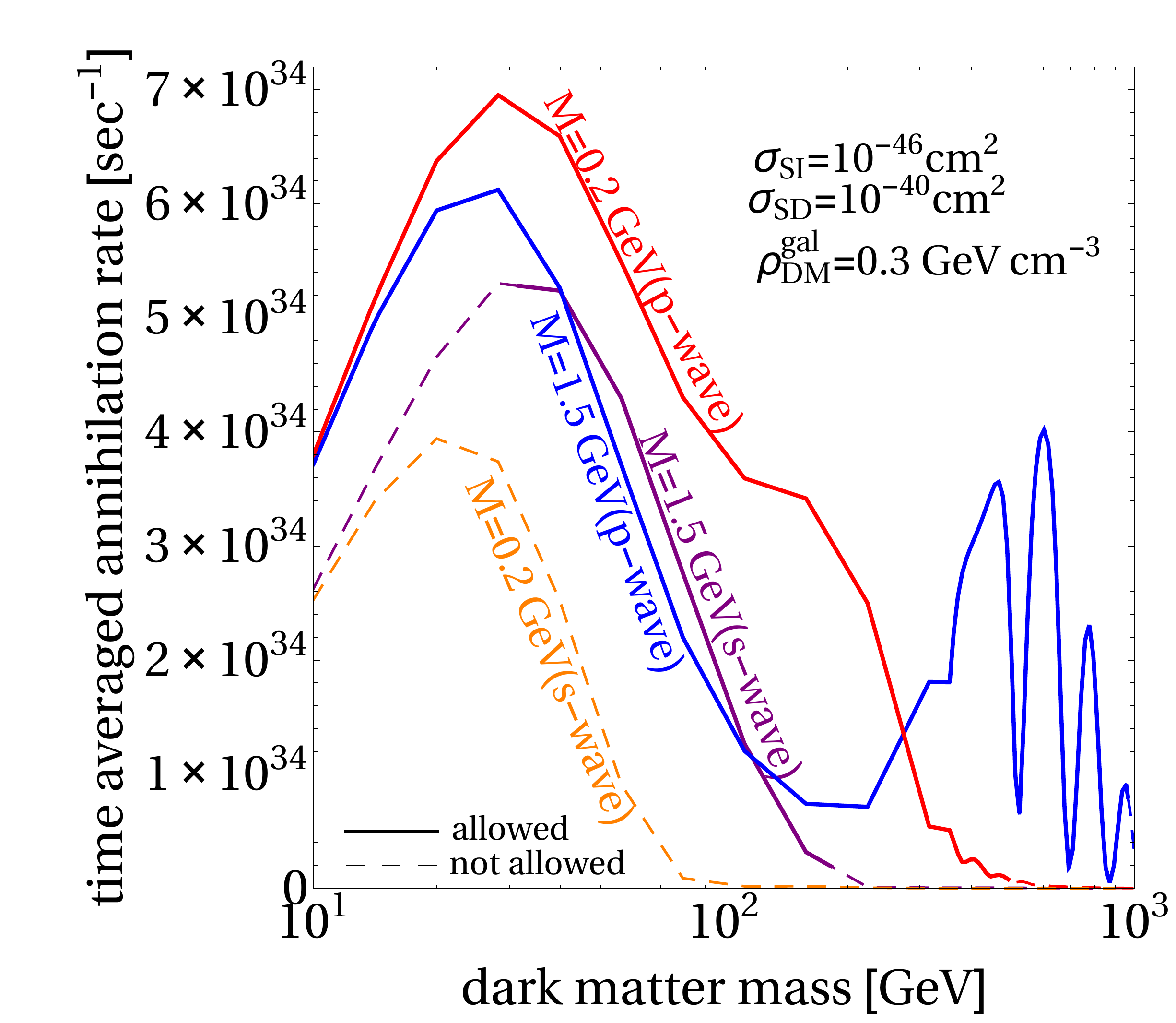}
  \caption{Average DM annihilation rate during the annihilation burst
    accompanying a supernova (duration $\Delta t_\text{burst}$).
    Dashed lines indicate parameter regions that
    are in tension with CMB constraints~\cite{Slatyer:2015jla, An:2016kie}.
    }
  \label{fig:burst-rate}
\end{figure}

Note that the annihilation rate during the burst depends on the location
of the star in the galaxy through $\rho_\text{DM}^\text{gal}$.
For a star located about 1\,kpc from Earth in the direction of the galactic center,
it is plotted in \cref{fig:burst-rate}.  We
see that, with increasing $m_\text{DM}$, the intensity of the burst increases at
first thanks to the increase in $C_\text{ann}^\text{SN}$, cf.\
\cref{eq:C-ann-SN}. At $m_\text{DM} > \text{few} \times 10$\,GeV, the intensity
decreases again because of the lower galactic DM density, because
capture of particles much heavier than the nuclei in
the star is kinematically inefficient, and because of
Sommerfeld-enhanced annihilation already at early times. The
wiggles in the blue and red curves at $m_\text{DM} \gtrsim 300$\,GeV are due to
Sommerfeld enhancement.

\begin{figure}
  \centering
  \includegraphics[width=0.9\columnwidth]{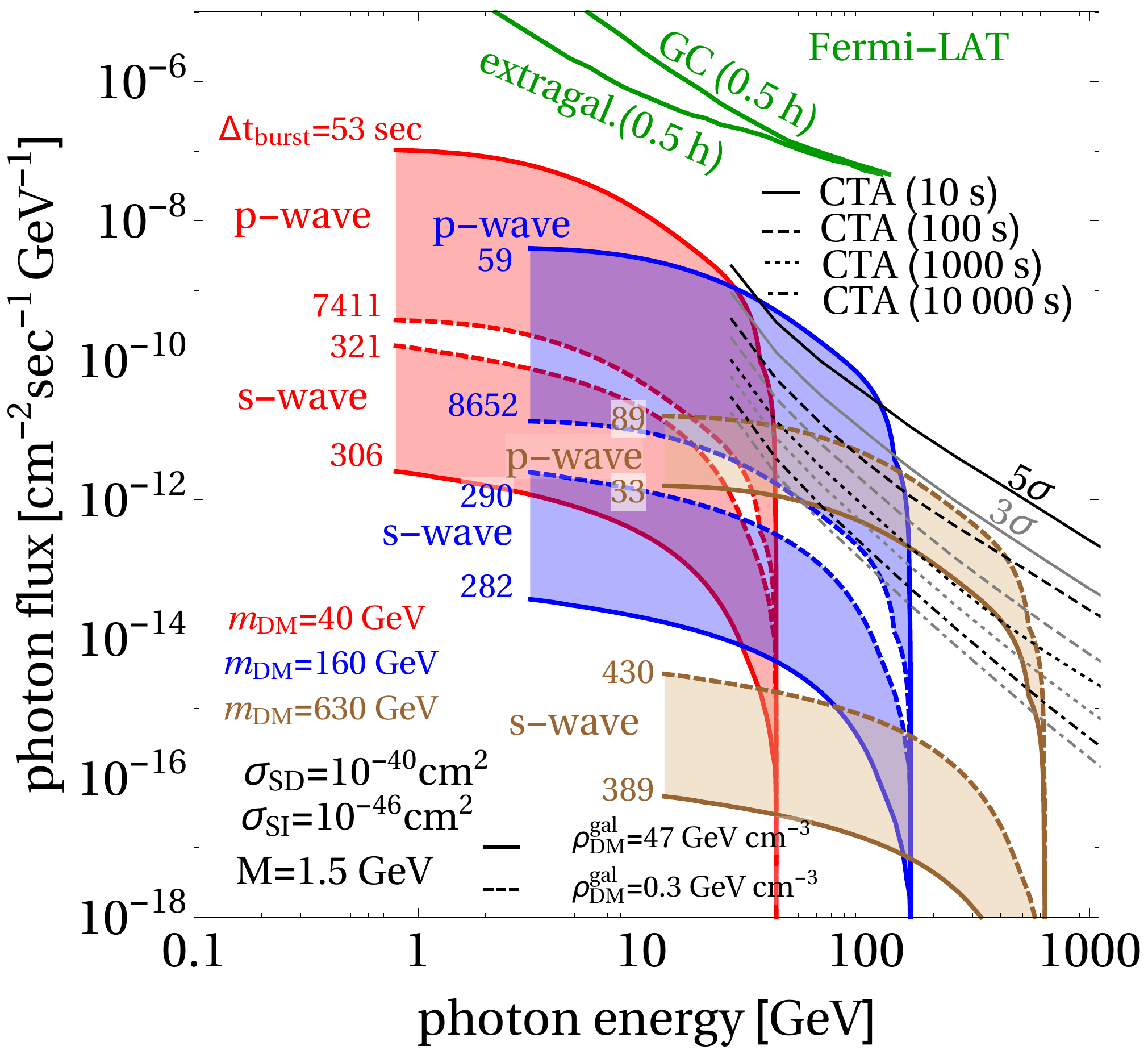}
  \caption{Expected photon flux from a ``dark gamma ray burst'',
    assuming DM annihilates into dark photons ($s$-wave) or dark scalars
    ($p$-wave) of mass $M$, which decay to SM particles after leaving the
    supernova, but before reaching Earth.  Results are shown for
    a star 0.1\,kpc (solid) or 7\,kpc (dashed) from the Galactic Center,
    and for various DM masses (different colors).
    We compare to the
    point source sensitivities of Fermi-LAT and CTA~\cite{Bernlohr:2012we,
    Hinton:2013}.
    }
  \label{fig:photon-spectrum}
\end{figure}

{\bf Dark gamma ray bursts.}
If the DM annihilation products are
SM particles, they cannot escape the newly born neutron star and
are therefore unobservable. This is true even for neutrinos.
Phenomenological prospects are brighter in the benchmark models discussed above, with DM
annihilating to dark photons $A'$ or dark scalars $\phi$~\cite{Batell:2009zp,
Schuster:2009fc, Meade:2009mu, Bell:2011sn, Feng:2016ijc}.
Dark photons can decay to SM particles through a kinetic
mixing interaction of the form $\mathcal{L}_\text{mix} = \epsilon F_{\mu\nu}
F'^{\mu\nu}$, where $F_{\mu\nu}$ ($F'_{\mu\nu}$) is the QED ($U(1)'$)
field strength tensor.  Similarly, dark scalars $\phi$ could decay through a
Higgs portal coupling of the form $\lambda \ev{\phi} \phi (H^\dag H)$, where
$H$ is the SM Higgs doublet.  We assume the coupling constant $\epsilon$ or
$\lambda \ev{\phi}$ to be such that the decay length of $A'$ or $\phi$ is much
larger than the radius of the star, but much smaller than its distance from
Earth. Among the secondary particles produced in the decays will be photons
from final state radiation or from meson decays.  These provide an
observable signal, which we dub a ``dark gamma ray burst''. The energy
distribution $dN_\gamma/dE_\gamma$ of burst photons is found along the
lines of ref.~\cite{Liu:2014cma} and plotted in \cref{fig:photon-spectrum}.  In
the plot, we also compare to the point source sensitivities of the Fermi Large
Area Telescope (Fermi-LAT) and of the future \v{C}erenkov Telescope Array
(CTA)~\cite{Bernlohr:2012we,Hinton:2013}.
For the latter, we show a conservative estimate ($5\sigma$ excess and $\geq 10$
signal events per bin) and a more optimistic one ($3\sigma$ excess and $\geq 2$
signal events per bin). Taking into account that the dark gamma ray burst signal is spread over
several energy bins, we find that the topmost curves for $m_\text{DM} =
40,\,160,\ \text{and}\ 630$\,GeV are detectable at the
$9.8\sigma$, $41.6\sigma$, and $15.6\sigma$ levels, respectively. For
$m_\text{DM} = 160$\,GeV and p-wave annihilation, a dark gamma ray burst
at 7\,kpc from the galactic center can be detectable at $6.4\sigma$.
We conclude that, for a sizeable range of DM masses, CTA
may hope to observe dark gamma ray bursts if annihilation is a $p$-wave
process and if the supernova occurs during the 10\% duty cycle of CTA.
The fact that $p$-wave annihilation is easier to observe than
$s$-wave annihilation sets dark gamma ray bursts apart from other indirect DM
signatures. It is also interesting to observe that, for $p$-wave annihilation
of not too heavy DM particles,
the expected signal from a star close to the Galactic Center (GC) is
\emph{larger} than for a nearby star because of the strong dependence of
$N_0$ and $\Delta t_\text{burst}$ on $\rho_\text{DM}^\text{gal}$ in $p$-wave models.

\begin{figure}
  \centering
  \includegraphics[width=0.7\columnwidth]{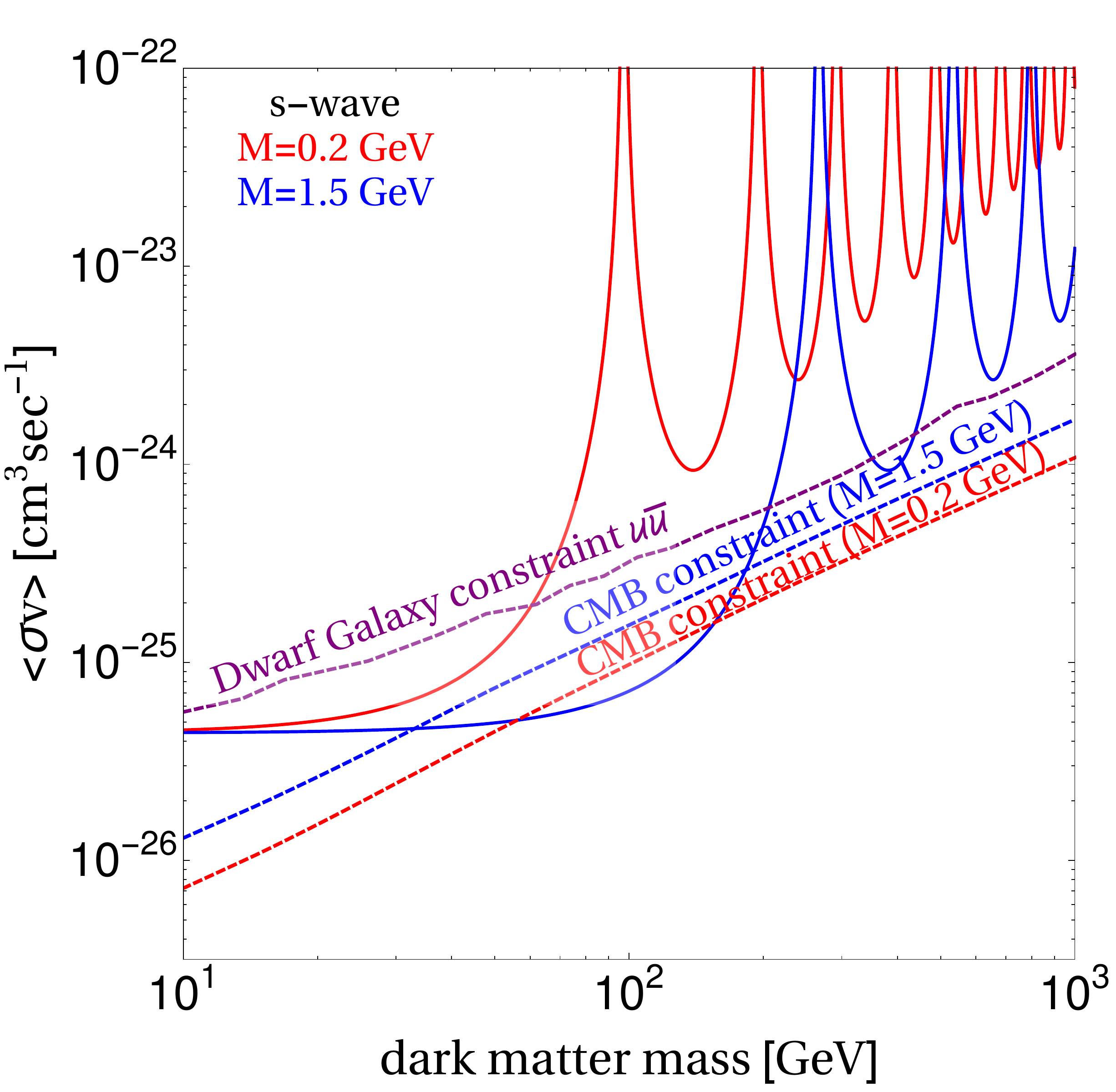}
  \caption{Predicted DM annihilation cross section for models with $s$-wave
    annihilation into a light force mediators of mass $M$ (solid curves),
    compared to constraints from CMB observations~\cite{Slatyer:2015jla} and
    from gamma ray observations of dwarf galaxies~\cite{Ackermann:2013yva}.
    Note the strong impact of Sommerfeld enhancement on the model prediction
    (peaks and dips at $m_\text{DM} \gtrsim 100$\,GeV).
    }
  \label{fig:constraints}
\end{figure}

{\bf Constraints.}
Models with a light force mediator $A'$ or $\phi$ are constrained
by other indirect DM searches.  Due to Sommerfeld enhancement,
DM annihilation is particularly
strong in environments with low DM velocities, therefore the strongest
constraints come from the Cosmic Microwave Background
(CMB) \cite{Slatyer:2015jla} and from gamma ray observations of dwarf galaxies
\cite{Ackermann:2013yva}. For $s$-wave annihilation, these constraints are shown
in \cref{fig:constraints}. If $M = 1.5$\,GeV, only DM masses between $\sim 30$\,GeV
and $\sim 200$\,GeV are allowed, while for $M = 0.2$\,GeV, no open parameter
space remains.  For $p$-wave annihilation (not shown in \cref{fig:constraints}),
Sommerfeld enhancement at low velocity is counteracted by smaller tree level
cross sections. Therefore, in the $p$-wave case, all DM masses are allowed
if the annihilation cross section at freeze-out is at the level required to
explain the DM abundance today and if $M \gtrsim 1$\,GeV. For lighter mediators,
bound state formation may lead to stronger limits~\cite{An:2016kie}.


{\bf Summary.}
We computed the properties of the DM core of a massive star throughout its
life, concluding that the death of the star in a supernova explosion is
accompanied by an intense DM annihilation burst.  If the DM annihilation
products are able to leave the exploding star and decay to SM particles later,
this may lead to a burst of secondary photons (``dark gamma ray burst''),
potentially observable in CTA.


{\bf Acknowledgements.}
We are indebted to Alexander Heger for providing the data from \cite{Heger:2002cn,
Langer:1997bi, Woosley:2007as, Woosley:2006gw} in
machine-readable form and for discussing it with us. We are also grateful to
Jatan Buch for useful discussions. The authors have
received funding from the German Research Foundation (DFG) under Grant Nos.\
\mbox{KO~4820/1--1}, FOR~2239 and GRK~1581, and from the European Research
Council (ERC) under the European Union's Horizon 2020 research and innovation
programme (grant agreement No.\ 637506, ``$\nu$Directions'').


\bibliography{dark-nova}

\end{document}